\documentclass[pre,amssymb,superscriptaddress,showpacs,floatfix,a4paper,twocolumn]{revtex4-1}

\usepackage{graphicx}% Include figure files
\usepackage{dcolumn}% Align table columns on decimal point
\usepackage{bm}% bold math
\usepackage{amsmath}
\usepackage{amssymb}
\usepackage{xcolor,cancel}

\begin{document}

\title{Numerical integration of the stochastic Landau-Lifshitz-Gilbert
equation in  generic time-discretization schemes}

\author{Federico Rom\'a}
\affiliation{Departamento de F\'{\i}sica,  Universidad Nacional de San Luis \& INFAP CONICET, 
Chacabuco 917, D5700BWS San Luis, Argentina}
\author{Leticia F. Cugliandolo}
\affiliation{Sorbonnes Universit\'es, Universit\'e Pierre et Marie Curie - Paris 6, 
Laboratoire de Physique Th\'eorique et Hautes \'Energies UMR 7589, 4 Place Jussieu, 
Tour 13, 5\`eme \'etage, 75252 Paris Cedex 05, France}
\author{Gustavo S. Lozano}
\affiliation{Departamento de F\'{\i}sica, FCEYN Universidad de Buenos Aires \& IFIBA CONICET, 
Pabell\'on 1 Ciudad Universitaria, 1428 Buenos Aires, Argentina}

\begin{abstract}
We introduce a numerical method to integrate the
stochastic Lan\-dau-Lifshitz-Gilbert equation in spherical
coordinates for generic discretization schemes. This method conserves the
magnetization modulus and ensures the approach to equilibrium under
the expected conditions.
We test the algorithm on a benchmark problem: the dynamics of a uniformly magnetized
ellipsoid. We investigate the influence of various parameters, and in particular, we analyze the
efficiency of the numerical integration, in terms of the number of steps needed to reach a
chosen long-time with a given
accuracy.
\end{abstract}

\maketitle

\section{Introduction}

The design of magnetic devices used to store and process information
crucially relies on a detailed understanding of how the magnetization dynamics
are influenced not only by external magnetic fields but also
by dissipation and thermal fluctuations~\cite{Hillebrands2002,Bertotti09}. In the simplest
scenario, the time evolution of the magnetization is governed by the stochastic
generalization of the Landau-Lifshitz-Gilbert (LLG) equation introduced by Brown
to study the relaxation of ferromagnetic nanoparticles~\cite{Brown}. In recent years,
much attention has been directed to the theoretical and experimental
understanding of how the magnetization can be manipulated with spin polarized
currents via the spin torque effect originally discussed by
Slonczewski~\cite{Slonczewski1996} and Berger~\cite{Berger1996}, an effect that can be described
by a simple generalization of this equation.

Explicit analytical solutions to the stochastic LLG
equation are available in very few cases; in more
general circumstances information has to be obtained by direct numerical
simulation of the stochastic equation, the study of the associated Fokker-Planck
equation (see Ref.~\onlinecite{review} for a recent review), or via functional methods~\cite{Aron14}.

The stochastic LLG  equation is a stochastic equation
with {\em multiplicative} noise. It is a well-known fact that in these cases, a careful analysis of
the stochastic integration prescriptions is needed to preserve the physical properties of the model.
In the stochastic LLG case  one should force the modulus of the magnetization 
to stay constant during evolution,
and different schemes (Ito, Stratonovich or the generic `alpha' prescription) require the addition 
of different drift terms
to preserve this property (for a recent discussion see Ref.~\onlinecite{Aron14}).  
All these issues are by now well-understood and they are also easy to 
implement in the continuous time treatment of the problem. Nevertheless,
this problem has not been analyzed in as much detail in the numerical formulation of the equation.

Indeed, most of the works focusing on the numerical analysis of the stochastic
equation use Cartesian coordinates \cite{Garcia98,Nowak2000,Li2004,Cheng2005,Cheng2006,dAquino,Usadel2006,Kazantseva2008,Titov2010,Weiler2011,Haase2012}.
Although there is nothing
fundamentally wrong with this coordinate system, most algorithms based on it
do not preserve, in an automatic way, the norm of the magnetization during
time evolution.
These algorithms require the explicit
magnetization normalization after every time step, a trick that is often
hidden behind other technical difficulties \cite{Martinez2004,Cimrak2007}.
This problem can be avoided only if the specific
midpoint prescription (Stratonovich) is used~\cite{dAquino}.

Given that the modulus of the magnetization should be constant by construction, a more
convenient way to describe the time evolution should be to use the spherical
coordinate system. Despite its naturalness, no detailed analysis of this
case exists in the literature.  The aim of this work is to
present a numerical algorithm to solve the LLG equation in the spherical coordinates system and
to discuss in detail how different discretization prescriptions are related,
an issue which is not trivial due to the multiplicative character of
the thermal noise.

In order to make precise statements,
we focus on the study of
the low-temperature dynamics of an ellipsoidal Cobalt nanoparticle,
a system that has been previously studied in great detail by other groups~\cite{dAquino}.
Our goal is to introduce the numerical method in the simplest possible setting
and the uniaxial symmetric potential involved in this problem seems to us a very good choice.

The paper is organized as follows. In Sec.~\ref{sec:problem} we present the problem. We
first recall the stochastic LLG equation in Cartesian and spherical coordinates. 
In both cases we
discuss the drift term needed to ensure the conservation of the magnetization modulus
as well as the approach to Boltzmann equilibrium. We then describe the concrete problem that we solve
numerically. In Sec.~\ref{sec:numerical} we present the numerical analysis. We first introduce the
algorithm and then discuss the results. Section~\ref{sec:conclusions} is devoted to the conclusions.

\section{The problem}
\label{sec:problem}

\subsection{The stochastic Landau-Lifshitz-Gilbert equation}
\label{sec:sLLG}

The stochastic Landau-Lifshitz-Gilbert (sLLG) equation in the Landau formulation of 
dissipation~\cite{Landau} reads
\begin{eqnarray}
{\rm d}_t {\mathbf{M}} &=&
 - \frac{\gamma_0}{1+\gamma_0^2 \eta^2}
 \
\mathbf{M} 
\nonumber\\
&& \wedge
\left(
 {\mathbf{H}}_{\rm eff} + {\mathbf H}
 + \frac{\eta \gamma_0}{M_s} \ {\mathbf{M}} \wedge  ({\mathbf{H}}_{\rm eff} + {\mathbf H})
 \right) ,
 \label{eq:landau}
\end{eqnarray}
where ${\rm d}_t \equiv {\rm d}/{\rm d} t$.
$\gamma_0\equiv\gamma \mu_0$ is the product of $\gamma$, the gyromagnetic
ratio relating the magnetization to the angular momentum,
and $\mu_0$, the vacuum permeability constant.
The gyromagnetic factor is given by $\gamma=\mu_B g/\hbar$
 and in our convention $\gamma>0$  with
$\mu_B$  Bohr's magneton and $g$ Lande's $g$-factor.
The symbol $\wedge$ denotes a vector product.
For ${\mathbf H}=0$ the first term in the right-hand-side describes the magnetization
precession around the local effective magnetic field $\mathbf{H}_{\rm eff}$.
The term proportional to ${\mathbf M} \wedge ({\mathbf M} \wedge {\mathbf H}_{\rm eff})$
is responsible for dissipation. Thermal effects are introduced \`a la Brown via the random 
field ${\mathbf H}$~\cite{Brown}
which is assumed to be Gaussian distributed
with average and correlations
\begin{equation} \label{eq:HhatD}
\langle H_i(t) \rangle_{\mathbf H} = 0
\;, \ \ \
\langle  H_i(t)  H_j(t')  \rangle_{\mathbf H}  = 2 D \delta_{ij}  \delta(t-t')
\; ,
\end{equation}
for all $i,j=x,y,z$. The parameter $D$ is, for the moment, free
and is determined  below.  $\eta$ is the dissipation coefficient
and in most relevant physical applications,  $\gamma_0\eta\ll 1$. An equivalent way of
introducing dissipation was proposed by Gilbert~\cite{Gilbert} but we
have chosen to work with the Landau formalism in this work.

This equation conserves the modulus of ${\mathbf M}$ and takes to Boltzmann
equilibrium {\it only if} the Stratonovich, mid-point prescription, stochastic calculus is used.
Otherwise, for other stochastic discretization prescriptions, none of these physically expected
properties are ensured. The addition of a carefully chosen drift term is needed to recover
the validity of these  properties when other stochastic
calculi are used. The generic modified sLLG equation~\cite{Aron14}
\begin{eqnarray}
{\rm D}^{(\alpha)}_t {\mathbf{M}} &=&
 - \frac{\gamma_0}{1+\gamma_0^2 \eta^2}
 \
\mathbf{M} \wedge
\nonumber\\
&&
\left(
 {\mathbf{H}}_{\rm eff} + {\mathbf H}
 + \frac{\eta \gamma_0}{M_s} \ {\mathbf{M}} \wedge  ({\mathbf{H}}_{\rm eff} + {\mathbf H})
 \right)
 \; ,
\label{eq.LLG-covariant}
\end{eqnarray}
where the time-derivative has been replaced by the $\alpha$-covariant derivative
\begin{equation}
{\rm D}^{(\alpha)}_t= {\rm d}_t + 2D (1-2\alpha) \frac{\gamma_0^2}{1+\eta^2\gamma_0^2}
\; ,
\label{eq.covariant}
\end{equation}
ensures the conservation of the magnetization modulus and convergence to 
Boltzmann equilibrium for any value of $\alpha$. 
The reason for the
need of an extra term in the covariant derivative is that the chain-rule for time-derivatives 
of functions of the stochastic variable is not the 
usual one when generic
stochastic calculus is used. It involves an additional term 
(for a detailed explanation see Ref.~\onlinecite{Aron14}).
In addition, having modified the stochastic equation in this way, one easily proves that
the associated Fokker-Planck equation is independent of $\alpha$ and
takes the magnetization to its equilibrium Boltzmann distribution
at temperature $T$ provided the parameter $D$
is given by
\begin{equation}
D = \frac{\eta k_BT}{M_s V \mu_0}
\; ,
\end{equation}
where
$V$ is the volume of the sample that behaves as a single macrospin, $k_B$ the Boltzmann constant, 
and $M_s$ the saturation magnetization. The parameter $\alpha$ is constrained 
to vary in $[0, 1]$. The most popular conventions are the
Ito one that corresponds to $\alpha=0$ and the Stratonovich calculus which is defined by 
$\alpha=1/2$. 
Note that this is not in contradiction with the 
claim by Garc\'{\i}a-Palacios~\cite{Garcia-Palacios}
that Ito calculus does not take {\it his} LLG equation to Boltzmann equilibrium as he keeps,
as the starting point,  the same form for the Ito and Stratonovich calculations
and, consequently, he obtains the  Boltzmann result only for Stratonovich rules.

As the modulus of the magnetization is conserved, this problem admits a more natural representation in spherical coordinates.
The vector $\mathbf{M}$ defines the usual local basis
($\mathbf{e}_r, \mathbf{e}_\theta, \mathbf{e}_\phi$)
with
\begin{eqnarray}
\mathbf{M}(M_s,\theta,\phi) &\equiv&
%M_\mu \ \mathbf{e}_\mu(\theta,\phi) =
M_s\, \mathbf{e}_r(\theta,\phi)
\end{eqnarray}
and
\begin{eqnarray}
M_x(t) &=& M_s \sin \theta(t) \sin \phi(t)
\; ,
\nonumber\\
M_y(t) &=& M_s \sin\theta(t) \cos\phi(t)
\; ,
\label{sphtrans}
\\
M_z(t) &=& M_s \cos \theta(t)
\; .
\nonumber
\end{eqnarray}
The sLLG equation in this system of coordinates becomes~\cite{Aron14}
\begin{eqnarray}
{\rm d}_t M_s &=& 0
\; ,
\\
{\rm d}_t\theta
&=&
\frac{D(1-2\alpha)\gamma_0^2}{1+\eta^2\gamma_0^2}
\ \cot\theta
\nonumber\\
&&
+
\frac{\gamma_0}{1+\eta^2\gamma_0^2}
\left[ H_{{\rm eff},\phi} + H_\phi
\right.
\nonumber\\
&&
\qquad\qquad\qquad
\left.
+ \eta\gamma_0 ( H_{{\rm eff},\theta} + H_\theta) \right]
\;,
\label{eq:spherical-Landau1}
\\
\sin\theta \ {\rm d}_t \phi
&=&
\frac{\gamma_0}{1+\eta^2\gamma_0^2}
\left[
\eta\gamma_0 ( H_{{\rm eff},\phi} + H_\phi )
\right.
\nonumber\\
&&
\left.
\qquad\qquad\quad
- ( H_{{\rm eff},\theta} + H_\theta)
\right]
\; ,
\qquad
\label{eq:spherical-Landau2}
\end{eqnarray}
where the $\theta$ and $\phi$ components of the stochastic field
are defined as
\begin{eqnarray}
H_\theta &=&
H_x \cos\theta \cos\phi + H_y \cos\theta \sin\phi - H_z \sin\theta
\; ,
\label{eq:def-Htheta}
\\
H_\phi &=&
-H_x \sin\phi + H_y \cos\phi
\label{eq:def-Hphi}
\; ,
\end{eqnarray}
and similarly for ${\mathbf H}_{\rm eff}$.

We introduce an adimensional time, $\tau=\gamma_0 M_s t$,
and the adimensional damping constant $\eta_0=\eta \gamma_0$,
and we normalize the field and the magnetization
by $M_s$ defining, ${\mathbf m}={\mathbf M}/M_s$,
${\mathbf h}_{\rm eff}={\mathbf H}_{\rm eff}/M_s$,
${\mathbf h}={\mathbf H}/M_s$, to write the equations as
\begin{eqnarray}
{\rm d}_\tau\theta
&=&
\frac{D\gamma_0(1-2\alpha)\gamma_0}{M_s (1+\eta_0^2)}
\ \cot\theta
\nonumber\\
&&
+
\
\frac{1}{1+\eta_0^2}
\left[ h_{{\rm eff},\phi} + h_\phi
+ \eta_0 ( h_{{\rm eff},\theta} + h_\theta) \right]
,
\label{eq:spherical-Landau-adim}
\\
\sin\theta \ {\rm d}_\tau \phi
&=&
\frac{1}{1+\eta_0^2}
\left[
\eta_0 ( h_{{\rm eff},\phi} + h_\phi )
- ( h_{{\rm eff},\theta} + h_\theta)
\right]
 .
\qquad
\label{eq:spherical-Landau-adim2}
\end{eqnarray}
The random field statistics is now modified
to $\langle h_i (\tau) \rangle_{\mathbf h} = 0 $ and %\linebreak
$\langle h_i(\tau) h_j(\tau') \rangle_{\mathbf h}= 2D\gamma_0/M_s
\ \delta_{ij} \ \delta(\tau-\tau')$.

\subsection{The benchmark}
\label{sec:model}

We focus here on the dynamics of a uniformly magnetized ellipsoid with energy
per unit volume
\begin{equation}
U= -\mu_0 {\mathbf M} \cdot {\mathbf H}_{\rm ext}
+ \frac{\mu_0}{2} (d_x M_x^2 + d_y M_y^2 + d_z M_z^2)
\;.
\end{equation}
${\mathbf H}_{\rm ext}$ is the external magnetic field and $d_x, \ d_y, \ d_z$ are
the anisotropy parameters. This case has been analized in detail in Ref.~\onlinecite{dAquino} 
and is used as a benchmark with which to compare our results.
We normalize the energy density by $\mu_0 M^2_s$, and
write
\begin{equation}
u= -{\mathbf m} \cdot {\mathbf h}_{\rm ext}
+ \frac{1}{2} (d_x m_x^2 + d_y m_y^2 + d_z m_z^2)
\;.
\end{equation}
The effective magnetic field is ${\mathbf H}_{\rm eff}
= - \mu_0^{-1} \partial U/\partial {\mathbf M}$.
Once normalised by $M_s$, it reads
\begin{equation}
{\mathbf h}_{\rm eff} =
{\mathbf h}_{\rm ext}
- (d_x m_x {\mathbf e}_x + d_y m_y {\mathbf e}_y
+ d_z m_z {\mathbf e}_z)
\; .
\end{equation}

We study the dynamics of a Cobalt nanoparticle of prolate spherical form
with radii $c = 4$ [nm] (in the $z$ easy-axis direction) and $a = b = 2$~[nm] (in the $x$ and $y$ directions,
respectively), yielding a volume $V = 6.702 \times 10^{-26}$~[m$^3$].
There is no external applied field, the saturation magnetization is
$M_s = 1.42 \times 10^6$~[A/m],
the uniaxial anisotropy constant in the $z$ direction is $K_1 = 10^5$ [J/m$^3$],
and the temperature is $T = 300$ [K]. In the following we work with the
adimensional damping constant $\eta_0=\eta\gamma_0$
and the physical value for it is $\eta_0 = 0.005$.
For this nanoparticle one has $d_x = N_x= d_y = N_y = 0.4132$
(where $N_i$ are the demagnetization factors~\cite{Cullity}), and $d_z = N_z - 2 K_1/(\mu_0 M_s^2)
= 0.0946$ since $N_z=0.1736$.
The constant $\gamma_0$ takes the value $2.2128 \times 10^5$ [m/(As)].
We recall that in Ref.~\onlinecite{dAquino} the time-step used in the
numerical integration is $\Delta t = 1.6$ [ps], that is
equivalent to $\Delta \tau = (\gamma_0M_s)\Delta t = 0.5$.

For future reference, we mention here that in the absence of an external field, ${\mathbf h}_{\rm ext}=0$,
the energy density $u$ can be written in terms of the $z$ component of the magnetization as
\begin{equation}
u = \frac12 \left[ d_x (1-m_z^2) + d_z m_z^2\right]
\; .
\end{equation}
Note also that in this system the anisotropy-energy barrier is
$V \Delta U=\mu_0 M_s^2 V (d_x-d_z)/2$, and therefore the ratio
$k_B T /(V \Delta U) \approx 0.153$, that indicates that the dynamics
take place in the low-temperature regime.

\section{Numerical analysis}
\label{sec:numerical}

In this section we first give some details on the way in which we
implemented the numerical code that integrates the equations,
and we next present our results.

\subsection{Method}

First, we stress an important fact explained
in Ref.~\onlinecite{Aron14}: the random fields $h_\theta$ and $h_\phi$ are
not Gaussian white noises but acquire, due to the prefactors
that depend on the angles, a more complex distribution function.
Therefore, we do not draw these random numbers but the
original Cartesian components of the random field
which are uncorrelated Gaussian white noises. We then recover the
field $h_\theta$ and $h_\phi$ by using Eqs.~(\ref{eq:def-Htheta}) and (\ref{eq:def-Hphi})
and the time-discretization of the product explained below.
Most methods used to integrate the sLLG equation rely on explicit
schemes. Such are the cases of the Euler and Heun methods.
While the former converges to the Ito solution, the latter 
leads to the
Stratonovich limit \cite{Garcia98}.  To preserve the module of ${\mathbf
m}$, in these algorithms it is necessary to normalize the magnetization in
each step, a nonlinear modification of the original sLLG dynamics
\cite{Cimrak2007}.  Implicit schemes, on the other hand, are very stable
and, for example, the mid-point method (Stratonovich stochastic calculus)
provides a simple way to automatically preserve the module under
discretization \cite{dAquino}.  In what follows, we describe our
numerical-implicit scheme which keeps the module length constant and, unlike
previous approaches, is valid for any discretization
prescription.

Next, we define the $\alpha$-prescription angular variables
according to
\begin{eqnarray}
\theta^\alpha(\tau) \equiv \alpha \theta(\tau+\Delta\tau) + (1-\alpha)
\theta(\tau)
\; ,
\\
\phi^\alpha(\tau) \equiv \alpha \phi(\tau+\Delta\tau) + (1-\alpha)
\phi(\tau)
\; ,
\end{eqnarray}
with $0 \leq \alpha \leq 1$. In the following we use the short-hand notation $\theta^\alpha(\tau) =\theta^\alpha_\tau$,
$\theta(\tau) =\theta_\tau$, and so on. The discretized dynamic equations
now read $F_\theta=0$ and $F_\phi=0$ with
\begin{eqnarray}
F_\theta & \equiv  &
- \ (\theta_{\tau+\Delta \tau}-\theta_\tau)+
D_0 \Delta \tau \frac{(1-2\alpha)}{(1+\eta_0^2)} \cot\theta^\alpha_\tau
\nonumber \\
&&
+ \
\frac{\Delta \tau}{1+\eta_0^2} \left[ h_{{\rm eff},\phi}^\alpha + \eta_0 h_{{\rm eff},\theta}^\alpha \right]
\nonumber \\
&&
+ \
\frac{1}{1+\eta_0^2} \left[ \Delta W_\phi + \eta_0 \ \Delta W_\theta \right] \label{eq:first}
\; ,
\\
F_\phi & \equiv  &
- \
(\phi_{\tau+\Delta \tau}-\phi_\tau)
 \nonumber \\
&&
+ \ \frac{\Delta \tau}{1+\eta_0^2} \left[\frac{ \eta_0 \ h_{{\rm eff},\phi}^\alpha - h_{{\rm eff},\theta}^\alpha }{\sin \theta^\alpha_\tau}\right]
\nonumber \\
&&
+ \
\frac{1}{1+\eta_0^2} \left[\frac{ \eta_0 \ \Delta W_\phi - \Delta W_\theta }{\sin \theta^\alpha_\tau}\right]
\; ,
 \label{eq:second}
\end{eqnarray}
where $D_0 = D\gamma_0/M_s$, the effective fields at the
$\alpha$-point are $h^\alpha_{{\rm eff},\theta} \equiv h_{{\rm eff},\theta}(\theta_\tau^\alpha, \phi_\tau^\alpha)$
and $h^\alpha_{{\rm eff},\phi} \equiv h_{{\rm eff},\phi}(\theta_\tau^\alpha, \phi_\tau^\alpha)$, and
$\Delta W_\phi=h_\phi \Delta \tau$ and $\Delta W_\theta=h_\theta \Delta \tau$.
As we said above, we first draw the  Cartesian components of the  fields ($i = x, y, z$) as
\begin{equation}
\Delta W_i = h_i \Delta \tau =  \omega_i \sqrt{2 D_0 \Delta \tau} ,
\end{equation}
where the $\omega_i$ are Gaussian random numbers with mean zero and variance one,
and we then calculate $\Delta W_\phi$ and $\Delta W_\theta$ using Eqs.~(\ref{eq:def-Htheta})
and (\ref{eq:def-Hphi}).

The numerical integration of the discretised dynamics
consists in finding the roots of the coupled  system of equations
$F_\theta=0$ and $F_\phi=0$ with the left-hand sides given in 
Eqs.~(\ref{eq:first}) and (\ref{eq:second}). We used a Newton-Raphson
routine~\cite{Press} and we imposed that the quantity $F_\theta^2 + F_\phi^2$
be smaller than $10^{-10}$. To avoid singular behavior when the
magnetization gets too close to the $z$ axis, $\theta=0$ or $\theta=\pi$,
we apply in these cases a $\pi/2$ rotation of the coordinate system
around the $y$ axis.

All the results we present below, averages and distributions, have been computed
using $10^5$ independent runs.

\subsection{Results}

\subsubsection{Stratonovich calculus}

\begin{figure}[t]
\includegraphics[width=7cm,clip=true]{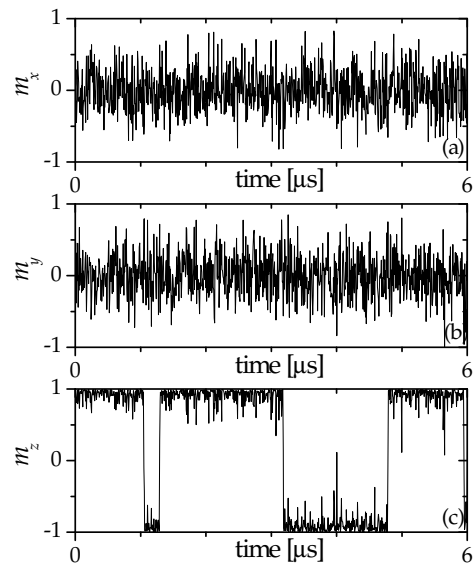}
\caption{A typical magnetization trajectory as a function of real time measured in $\mu$s
showing rapid fluctuations around
zero for $m_x$ (a) and $m_y$ (b) and telegraphic noise with sudden transitions between the
up and the down magnetization configurations for $m_z$ (${\mbox c}$). The initial condition is
${\mathbf m}=(0,0,1)$, $\alpha=0.5$, $\eta_0=0.005$,
$\Delta \tau = 0.5$, and $\tau_{max}= 2 \times 10^6$ that is equivalent to
$t_{max}=6.36$~$\mu$s. In this and all other figures the working temperature is $T=300$ K.}
\label{fig:uno}
\end{figure}

We start by using the Stratonovich discretization scheme,
$\alpha=0.5$, to numerically integrate the stochastic equation
using the parameters listed in Sec.~\ref{sec:model} which are the same
as the ones used in Ref.~\onlinecite{dAquino}. We simply stress here that these are
typical parameters (in particular, note the  small value of the
damping  coefficient $\eta_0$). Although we solved the problem in spherical coordinates, we illustrate
our results in Cartesian coordinates [using Eqs.~(\ref{sphtrans}) to transform back to
these coordinates] to allow
for easier comparison with the existing literature.

\paragraph{Trajectories.}
Figure~\ref{fig:uno} displays the three Cartesian components of the
magnetization, $m_x$, $m_y$, and $m_z$, as a function of time for a single
run starting from an initial condition that is perfectly polarized along
the $z$ axis, ${\mathbf m}=(0,0,1)$. The data show that while the $x$ and $y$ components
fluctuate around zero, the $z$ component has telegraphic noise, due to  the very
fast magnetization reversal from the `up' to the `down' position and
vice versa. Indeed, the working temperature we are using is rather low,
but sufficient to drive such transitions.

\begin{figure}[t]
\includegraphics[width=7cm,clip=true]{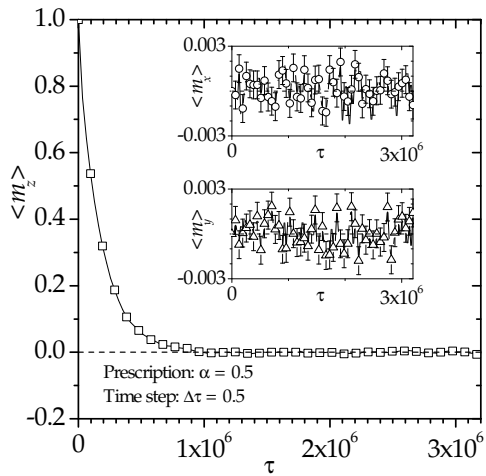}
\caption{Average value of the magnetization $z$ component as a function of $\tau$.
Insets: $\tau$ dependence of the other two components, $\langle m_x\rangle $ and $\langle m_y\rangle$.
$\alpha = 0.5$, $\eta_0 = 0.005$, and $\Delta\tau =0.5$.}
\label{fig:dos}
\end{figure}

\paragraph{Equilibrium criteria.}
In Fig.~\ref{fig:dos} we show the relaxation of the thermal average of the
$z$ component, $\langle m_z\rangle$, evolving from the totally polarized initial
condition, ${\mathbf m} = (0,0,1)$, during a maximum adimensional time $\tau_{max} = 3.2 \times 10^6$.
In the inset one can see temporal fluctuations around zero in the averages of the
other two components, $\langle m_x \rangle$ and $\langle m_y \rangle$. The error bars in these and other plots
are estimated as one standard deviation from the data average, and when these are smaller than the
data points we do not include them in the plots. The data in Fig.~\ref{fig:dos}
demonstrate that for times shorter than $10^6$ the system is still out of equilibrium
while for longer times this average is very close to the equilibrium expectation,
$\langle m_z\rangle_{\rm eq} = 0$.

\begin{figure}[t]
\includegraphics[width=7cm,clip=true]{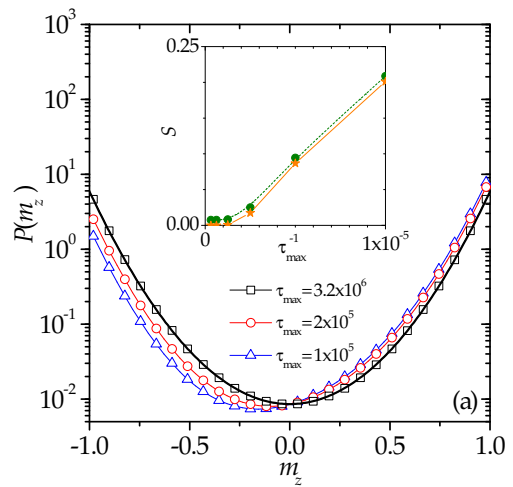}
\includegraphics[width=7cm,clip=true]{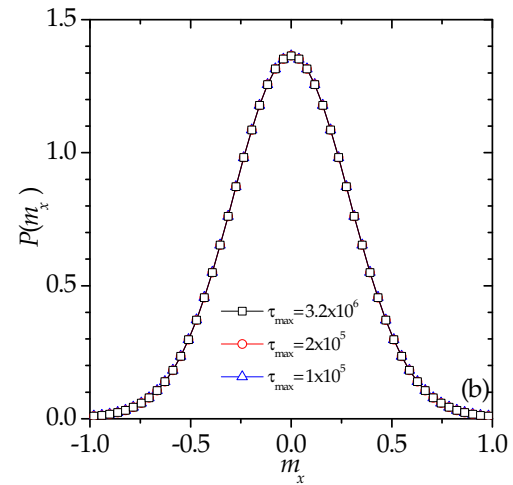}
\caption{(Color online)
(a) $P(m_z)$, on a linear-log scale, obtained as explained in the  text for
the three values of $\tau_{max}$ given in the key compared
to the exact equilibrium law (solid line). Inset: The parameter $S$
defined in Eq.~(\ref{eq:S-def}) as a function of ${\tau}^{-1}_{max}$. The upper (dotter) curve
was computed using the exact pdf $P_{\rm eq}(m_z)$, while the
lower (solid) one, which gets closer to zero, was computed using a finite number of bins to approximate
the exact $P_{\rm eq}(m_z)$.
(b) $P(m_x)$,  on a double-linear scale, for the same runs.
$\alpha = 0.5$, $\eta_0 = 0.005$, and $\Delta\tau =0.5$.}
\label{fig:tres}
\end{figure}

\begin{figure}[t]
\includegraphics[width=7cm,clip=true]{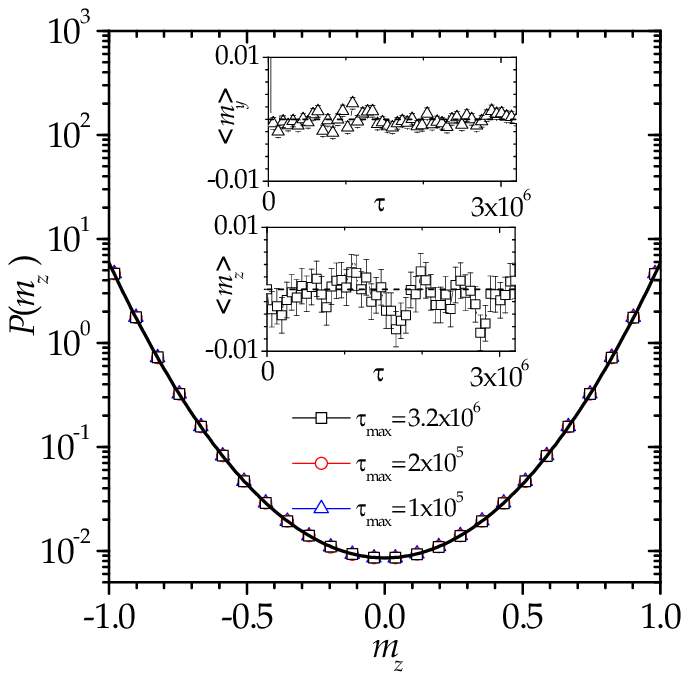}
\caption{(Color online)
 $P(m_z)$, on a linear-log scale, for three values of $\tau_{max}$ given in the key,
compared to the exact equilibrium law (solid line).
Inset: Averages of $\langle m_y\rangle$ and $\langle m_z\rangle$
as a function of $\tau$. The parameters are the same as in Fig.~\ref{fig:dos}
but the initial condition is ${\mathbf m} = (0, 1, 0)$.
}
\label{fig:cuatro}
\end{figure}

A more stringent test of equilibration is given by the analysis of the probability
distribution function (pdf) of the three Cartesian components $m_x$, $m_y$, and $m_z$. In Fig.~\ref{fig:tres} (a)
we present the numerical pdf's of $m_z$, $P(m_z)$, and we compare the numerical data to the theoretical distribution
function in equilibrium.  We computed the former
by sampling over the second half of the temporal window,
that is, by constructing the histogram with data collected over $\tau_{max}/2 \leq \tau
\leq \tau_{max}$, and then averaging the histograms over $10^5$ independent runs.
For the equilibrium $P_{\rm eq}(m_z)$ we note that the equilibrium probability density of the spherical angles is
\begin{equation}
P_{\rm eq}(\theta, \phi) \ d\theta d\phi \propto \frac{1}{4\pi} \ \sin\theta \ e^{-\varepsilon u} \ d\theta d\phi
\; ,
\end{equation}
where $\varepsilon=\mu_0 M_s^2 V/(k_B T)$, which implies
\begin{eqnarray}
P_{\rm eq}(m_z) \ dm_z &=&  -\frac12 \ e^{-\varepsilon u} \ dm_z \nonumber \\
&=& \frac{1}{Z} \ e^{-\frac{\varepsilon}{2}  \left[ d_x (1-m_z^2) + d_z m_z^2\right]}  \ dm_z
\; .
\label{eq:equil-pdf-mz}
\end{eqnarray}
Here $Z$ is the partition function and, for the parameters used in the simulation,
$\varepsilon=41$ and $Z=0.0244$.

It is quite clear from Fig.~\ref{fig:tres} that the numerical curves for the two shortest
$\tau_{max}$ are still far from the equilibrium one, having excessive weight on positive
values of $m_z$. The last curve, obtained for the longest running time, $\tau_{max}=3.2 \times
10^{6}$ is, on the contrary,
indistinguishable from the equilibrium one in this presentation. A more quantitative
comparison between numerical and analytic pdf's is given in the inset in Fig.~\ref{fig:tres}~(a),  where the
probability distribution `H-function'~\cite{Kubo92}
\begin{equation}
S(\tau_{max}) = \int_{-1}^1 dm_z \ P(m_z, \tau_{max}) \ \ln \frac{P(m_z, \tau_{max})}{P_{\rm eq}(m_z)}
\; ,
\label{eq:S-def}
\end{equation}
with $P_{\rm eq}(m_z)$ given in Eq.~(\ref{eq:equil-pdf-mz}),
is plotted  as a function of the inverse time $\tau_{max}^{-1}$. The two sets of data in the
inset correspond to $S$ computed with the continuous analytic form (\ref{eq:equil-pdf-mz}),
data falling above,
and with a discretized version of it, where the same number of bins as in the
numerical simulation is used (specifically, 51), and data falling below and getting very close to
zero for the longest $\tau_{max}$ used. The latter is the correct way of comparing analytic and numerical
data and yields, indeed, a better agreement with what was expected. Finally, in Fig.~\ref{fig:tres}~(b) we show the
pdf of $m_x$  for the same three $\tau_{max}$ used in Fig.~\ref{fig:tres}~(a), and
we observe a faster convergence to an equilibrium distribution with a form that is very close to a Gaussian.
For symmetry reasons the behavior of $m_y$ is the same.

Figure~\ref{fig:cuatro} shows the pdf's for the same set of parameters but starting from the
initial condition ${\mathbf m} = (0,1,0)$. The approach to equilibrium is faster in this case: all
curves fall on top of the theoretical one.
Insets show the time dependence of $\langle m_x\rangle $ and $\langle m_z\rangle $
which still fluctuate around zero with larger temporal fluctuations for the latter than the former.
We reckon here that the fluctuations of $\langle m_y\rangle $ and $\langle m_z\rangle$
are quite different. The oscillations of $\langle m_z\rangle$ around zero are due to the
telegraphic noise of this component and to the fact that the average is done over
a finite number of runs. The amplitudes of these oscillations tend to zero
with an increasing number of averages.

We conclude this analysis by stating that the dynamics in the spherical coordinate
system for the Stratonovich discretization scheme behave correctly, with the advantage of
keeping the norm of the magnetization fixed by definition.

\subsubsection{Generic calculus}

Although it was shown in Ref.~\onlinecite{Aron14} that in the $\Delta \tau \to 0$ 
limit every discretization of the stochastic
equation leads (at equilibrium) to the Boltzmann distribution, 
the numerical integration of the equations is
done at finite $\Delta \tau$ and then both, the time-dependent and the equilibrium 
averaged observables may depend on
$\Delta\tau > 0$. With this in mind, we investigated which discretization
scheme is more efficient in terms of computational effort. The aim of this
section is to study the $\Delta\tau$ dependence of the numerical results
for different values of $\alpha$ and to determine for which $\alpha$ one can get closer to the
continuous-time limit ($\Delta \tau \to 0$)  for larger values of $\Delta\tau$.

\begin{figure}[t]
\includegraphics[width=7cm,clip=true]{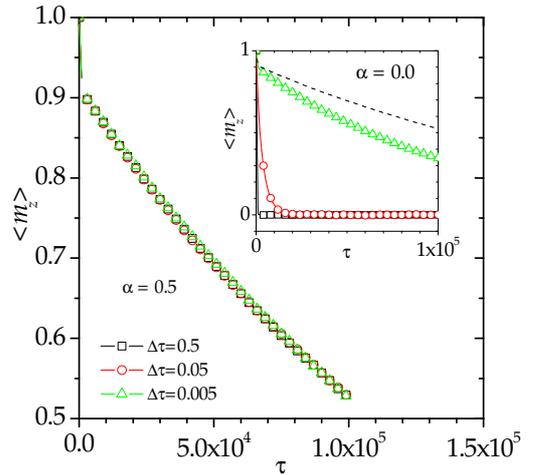}
\caption{
(Color online)
$\langle m_z\rangle$ as a function of $\tau$ for
 $\alpha=0.5$, $\eta_0=0.005$ and  $\Delta \tau=0.5$, $0.05$, and $0.005$.
Inset: $\langle m_z\rangle$ vs $\tau$ for
$\alpha=0.0$ and the same $\Delta \tau$ with the same symbol code as in the figure.
The dashed black line is a reference and corresponds to
 $\alpha=0.5$ and $\Delta \tau=0.005$. The initial condition is
 ${\mathbf m}=(0,0,1)$.
 }
 \label{fig:cinco}
\end{figure}

Figure~\ref{fig:cinco} shows the temporal dependence of the $\langle m_z \rangle$
for Stratonovich calculus, i.e.~for $\alpha=0.5$, for a small window of time and from
the initial condition ${\mathbf m} = (0,0,1)$. A very fast decay followed by a slow
relaxation is observed.  The phenomena can be well fitted by a sum of two exponential
functions, one (of small amplitude) describing the rapid intra-well processes and
the another one the dominant slow over-barrier thermo-activation~\cite{Garanin1996}.
Concretely, we used
\begin{displaymath}
\langle m_z\rangle (\tau)=A_1 \ e^{-\frac{\tau}{ \tau_1}}+ A_2 \ e^{-\frac{\tau}{ \tau_2}}
\end{displaymath}
and we found that the best description of data is given by
\begin{displaymath}
A_1=0.915 \; , \,\,\, \tau_1 = 1.8 \ 10^5 \;,  \,\,\, A_2=0.08  \; , \,\,\, \tau_2= 4.1 \ 10^2
\;  ,
\end{displaymath}
that is $A_1 \gg A_2$ and $\tau_1 \gg \tau_2$, consistently with statements in Ref.~\onlinecite{Garanin1996}.
In addition, we compared our result for $\tau_1$
to the one arising from Eq.~(3.6) in Ref.~\onlinecite{Garanin1996} for our parameters and we found
\begin{equation}
\tau^G_{1}=1.5 \ 10^5
\end{equation}
which is $15 \% $ less than our numerical estimate, a very reasonable agreement, in our opinion.

Most importantly, we reckon that the time-dependent results do not depend strongly on
$\Delta \tau$ for $\Delta\tau \leq 0.5$ (see Fig.~\ref{fig:cinco}, where
data for $\Delta\tau=0.005$, $0.05$, and $0.5$ prove
this claim) and we can assert that the master curve is as close as we can get, for the numerical accuracy we are
interested in, to the one for $\Delta\tau\to 0$,
that is, to the correct relaxation.
Instead, for other discretization prescriptions, the dependence
on $\Delta\tau$ is stronger. For example, for $\alpha=0$ (Ito calculus)
the curves for $\Delta\tau=0.05$ and $0.005$ are still notably
different from each other (see the insert to Fig.~\ref{fig:cinco}), and they have not yet converged to the physical
time-dependent average. Even smaller values of $\Delta \tau$ are needed to get close to the asymptotically
correct relaxation, shown by the dotted black line.
We do not show the pdf's here, but consistently, they are far away from the equilibrium one
for these values of $\Delta \tau$.

\begin{figure}[t]
\begin{center}
\includegraphics[width=7cm,clip=true]{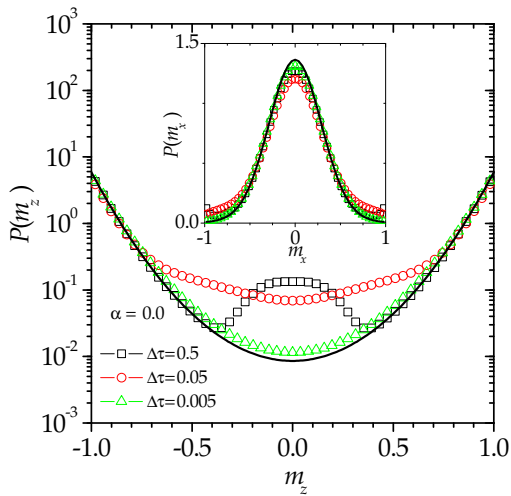}
\end{center}
\caption{
(Color online)
$P(m_z)$ for $\alpha=0.0$, $\eta_0=0.005$, and $\Delta \tau=0.5$, $0.05$ and $0.005$,
compared to the exact equilibrium law (solid line). The
initial condition is $\mathbf{m}=(0,1,0)$. Inset: Distributions $P(m_x)$ for the same runs and
using the same symbol code as in the rest of the figure,
compared to the limit (equilibrium) function shown in Fig.~\ref{fig:tres} (b) for $\alpha=0.5$ and
the longest $\tau_{max}$.}
\label{fig:seis}
\end{figure}

In Fig.~\ref{fig:seis} we use the  initial condition ${\mathbf m} = (0, 1, 0)$
to see whether the efficiency of the Ito calculus improves in this case. Although the values of
$\langle m_z\rangle $ are very close to the expected vanishing value both distributions, $P(m_z)$ (main panel)
and $P(m_x)$ (inset), are still far from equilibrium. We conclude that
also for this set of initial conditions smaller $\Delta \tau$ are
needed to reach the continuous-time limit. We have investigated other values of $\alpha\neq 0.5$ and
in all cases we have found that convergence is slower than for the $\alpha= 0.5$ case.

We conclude that the Stratonovich calculus is `more efficient' than all other
$\alpha$-prescriptions in the sense that one can safely use larger values of
$\Delta\tau$ (and therefore reach longer times) in the simulation.
This does not mean that other discretization schemes yield incorrect
results. For $\alpha\neq 0.5$ one must use smaller values of the time-step
$\Delta \tau$ to obtain the physical behavior.

\subsubsection{Effect of the damping coefficient}

\begin{figure}[t]
\begin{center}
\includegraphics[width=7cm,clip=true]{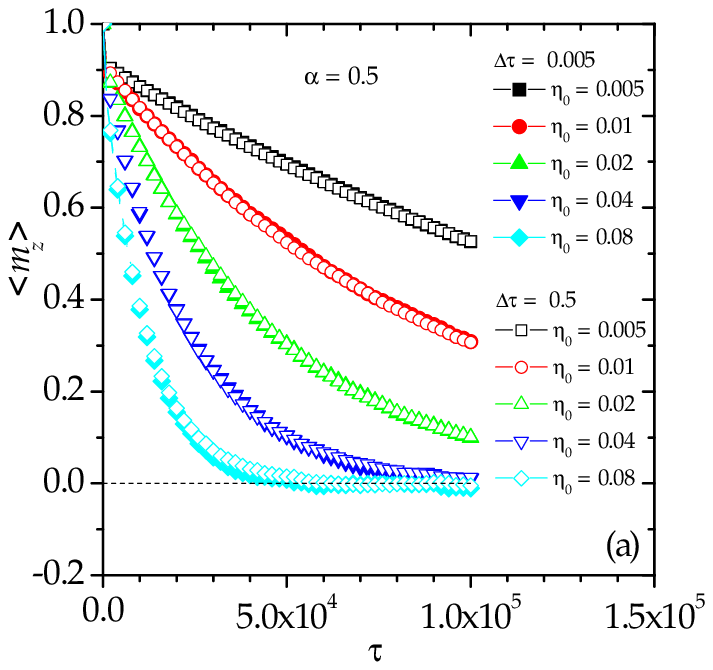}
\includegraphics[width=7cm,clip=true]{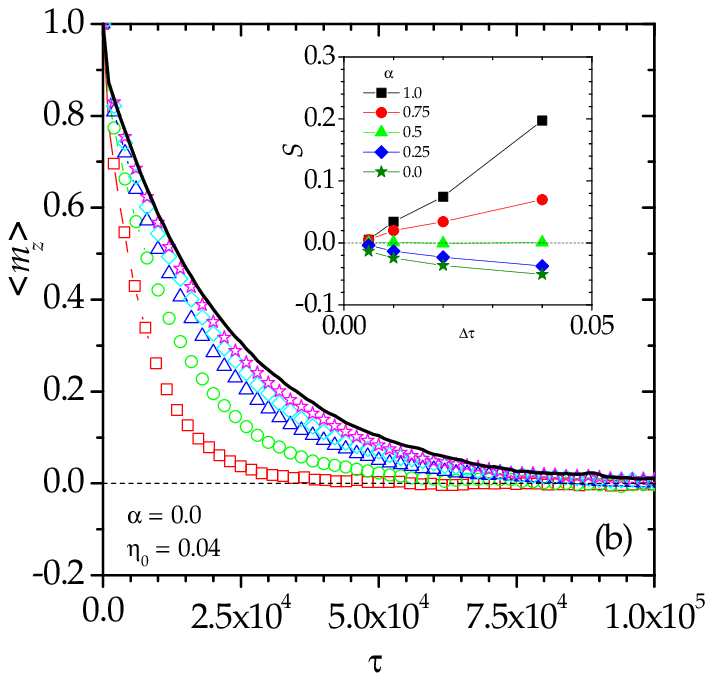}
\caption{
(Color online)
(a) $\langle m_z\rangle$ as a function of $\tau$ for
two values of the time increment, $\Delta \tau=0.5$ (open symbols) and $0.005$ (filled
symbols). $\alpha=0.5$ and several values of the damping coefficient $\eta_0$ (shown in
different colors) as defined in the legend. (b) Ito calculus, $\alpha=0$, and $\eta_0=0.04$.
Curves correspond to $\tau_{max}=10^5$ and different $\Delta\tau=0.005, \ 0.01, \ 0.02, \ 0.04, \ 0.08$ (from top to bottom).
The solid black line displays $\langle m_z\rangle$
for $\alpha=0.5$, $\eta_0=0.04$, and $\Delta \tau=0.005$. Inset:
The parameter $S$ defined in Eq.~(\ref{eq:S-def-dos}) for these curves, taking
as a reference the curve for $\alpha=0.5$.
}
\label{fig:siete}
\end{center}
\end{figure}

Previously, we studied the magnetization relaxation for a physically
small damping coefficient, $\eta_0=0.005$. Under these conditions
relaxation is very slow and it is difficult to reach convergence for
generic values of $\alpha$.  To overcome this problem, we increased
slightly the damping up to $\eta_0 = 0.08$. As a consequence, and because
we are still in the low damping regime, relaxation to equilibrium is expected to be faster
[in Fig.~\ref{fig:siete}~(a) we show below that this statement is correct].  Note that more subtle
issues can arise in the non axially symmetric case if initial
conditions are not properly chosen \cite{Kalmykov2010}. Then, in
this subsection we check whether the $\Delta\tau$ dependence found for
the $\alpha\neq 0.5$ calculus improves under these new dissipation
conditions.

In Fig.~\ref{fig:siete}~(a) we test the $\Delta\tau$ dependence of $\langle m_z\rangle$
for $\alpha=0.5$
and five values of $\eta_0$ ranging from $\eta_0=0.005$ to $\eta_0=0.08$
and increasing by a factor of two. Filled and open data points of the same color
correspond to $\Delta\tau=0.005$ and $\Delta\tau=0.5$, respectively. The agreement
between the two data sets is very good for all $\eta_0$.
Indeed the agreement is so good that the data are superimposed and it is hard to distinguish the different cases.
 The curves also show  that the dynamics are faster for increasing $\eta_0$.
 Figure~\ref{fig:siete}~(b) displays the decay of
$\langle m_z\rangle$ as a function of time for $\alpha=0$ and a rather large
value of the damping coefficient, $\eta_0=0.04$, for different time increments,
$\Delta\tau=0.005, \ 0.01, \ 0.02, \ 0.04, \ 0.08$. The curves tend to approach the reference one shown
by the solid black line and corresponding to $\alpha=0.5$ for decreasing values of
$\Delta\tau$. A quantitative measure of the convergence rate is given by another $S$
parameter, defined as
\begin{equation}
S(\Delta \tau,\alpha)= \frac{1}{\tau_{max}} \int_0^{\tau_{max}} d\tau \ \langle m_z\rangle_\alpha
\ \ln \left[ \frac{\langle m_z\rangle_\alpha}{\langle m_z\rangle_{0.5}}\right]
\; ,
\label{eq:S-def-dos}
\end{equation}
and shown in the inset for $\eta_0=0.04$.
Here, $\langle m_z\rangle_{0.5}$ is the average of $m_z$ for $\alpha=0.5$
and $\Delta \tau=0.005$, while $\langle m_z\rangle_\alpha$ is the curve corresponding to
other values of $\alpha$ and $\Delta \tau$.
For all $\alpha$-schemes $S$
tends to zero for $\Delta\tau\to 0$. Note that for $\alpha=0.5$ this parameter is
very close to zero for all the $\Delta\tau$ values shown in Fig.~\ref{fig:siete}~(b), confirmation
of the fact that this prescription yields very good results for relatively large
values of $\Delta\tau$ and it is therefore `more efficient' computationally.

\section{Conclusions}
\label{sec:conclusions}

In this paper we have introduced a numerical algorithm that solves the the sLLG dynamic equation
in the spherical coordinate system with no need for artificial normalization of the magnetization.
We checked that the algorithm yields the correct evolution
of a simple and well-documented problem~\cite{dAquino},
the dynamics of an ellipsoidal magnetic nanoparticle.
We applied the algorithm in the generic `alpha'-discretization prescription.
We showed explicitly  how the finite $\Delta \tau$ dynamics depend on $\alpha$, despite the fact that
the final equilibrium distribution is $\alpha$-independent.
We showed that at least for the case reported here, the Stratonovich mid-point prescription is the
`more efficient one' in the sense that the
dependence of the dynamics on the finite value of $\Delta \tau$
is less pronounced so, larger values of $\Delta \tau$
can be used to explore the long time dynamics.

We think it would be worthwhile to explore, both analytically and numerically,
if this is a generic result of the sLLG dynamics.
{\em A priori}, it is not clear what will be the optimal
prescription to deal with other problems such as a system under a non-zero longitudinal external
magnetic field  or for a more general non-axially symmetric potential
\cite{Coffey1995,Garanin1996,Kalmykov2010}.

Finally, we mention that it is well known that for a
particle on a line with multiplicative noise, the addition
of an inertial term acts as a regularization scheme that
after the zero mass limit ``selects'' the Stratonovich 
prescription (see Ref.~\onlinecite{Aron10}). For the case of
magnetization dynamics non Markovian generalizations
of the LLG equation have been considered in Refs.~\onlinecite{Miyasaki1998,Atxitia09}. 
It could be interesting to analyze in this case, how the markovian
limit relates to any specific stochastic prescription. 
We plan to report on this issue in the near future.

\begin{acknowledgments}
We thank C. Aron, D. Barci and Z. Gonz\'alez-Arenas for very helpful discussions on this topic.
F.R. acknowledges financial support from CONICET (Grand No. PIP 114-201001-00172) and Universidad
Nacional de San Luis, Argentina (Grand No. PROIPRO 31712) and thanks the LPTHE for hospitality during the
preparation of this work. L.F.C. and G.S.L. acknowledge financial support from FONCyT, 
Argentina (Grand No. PICT-2008-0516).
\end{acknowledgments}

\end{document}